\documentclass[aps,prd,showpacs,onecolumn,groupedaddress]{revtex4}
\usepackage{graphicx}
\usepackage{amsfonts}
\usepackage{amssymb}

\usepackage{ulem}
\usepackage{color}

\begin{document}

\title{Lattice QCD Study for Confinement in Hadrons}
\author{H.~Suganuma}
\affiliation{Department of Physics, Graduate School of Science, Kyoto University, \\
Kitashirakawaoiwake, Sakyo, Kyoto 606-8502, Japan}

\author{T.~Iritani}
\affiliation{Department of Physics, Graduate School of Science, Kyoto University, \\
Kitashirakawaoiwake, Sakyo, Kyoto 606-8502, Japan}

\author{F. Okiharu}
\affiliation{Faculty of Education, Niigata University, 
Ikarashi 2-8050, Niigata 950-2181, Japan}

\author{T.T. Takahashi}
\affiliation{Gunma National College of Technology, 
Maebashi, Gunma 371-8530, Japan}

\author{A.~Yamamoto}
\affiliation{Department of Physics, Graduate School of Science, Kyoto University, \\
Kitashirakawaoiwake, Sakyo, Kyoto 606-8502, Japan}

\date{\today}

\begin{abstract}

We study three subjects on quark confinement in hadrons 
in SU(3)$_{\rm c}$ lattice QCD.
From the accurate lattice calculation for more than 300 different patterns 
of three-quark (3Q) systems, we find that the static 3Q potential 
is well described by Y-Ansatz, i.e., 
the Coulomb plus Y-type linear potential. 
We also study the multi-quark (4Q, 5Q) potentials 
in lattice QCD, and find that they are well described 
by the one-gluon-exchange (OGE) Coulomb 
plus string-theoretical linear potential, 
which supports the {\it infrared string picture} 
even for the multi-quarks. 
The second subject is a lattice-QCD determination of 
the relevant gluonic momentum component for confinement. 
The string tension (confining force) is found to be almost unchanged 
even after cutting off the high-momentum gluon component above 1.5GeV 
in the Landau gauge.
In fact, {\it quark confinement originates from the low-momentum gluon 
below about 1.5GeV.} 
Finally, we consider a possible gauge of QCD 
for the quark potential model, 
by investigating ``instantaneous inter-quark potential''  
in generalized Landau gauge, which describes 
a continuous change from the Landau gauge to the Coulomb gauge.
\end{abstract}



\maketitle

\section{Introduction}

In 1966, Yoichiro Nambu \cite{N66} first proposed 
the SU(3)$_{\rm c}$ gauge theory, i.e., quantum chromodynamics (QCD),  
as a candidate for the fundamental theory of the strong interaction, 
just after the introduction of ``color'' \cite{HN65}.
Around 1970, he also proposed the string theory for hadrons \cite{N6970}, 
which describes hadron phenomenology.
In 1973, the asymptotic freedom of QCD was theoretically shown \cite{GWP73}, 
and QCD was established as the fundamental theory of 
the strong interaction.
However, in spite of its simple form, QCD creates thousands of hadrons 
and leads to various interesting nonperturbative phenomena 
such as color confinement and chiral symmetry breaking \cite{NJL61}.
Even now, it is very difficult to deal with QCD analytically 
due to its strong-coupling nature at the low-energy.
Instead, lattice QCD is now a reliable numerical method 
to analyze nonperturbative QCD. 
In this paper, we study three subjects on quark confinement 
in hadrons in SU(3)$_{\rm c}$ lattice QCD.

\section{Infrared String Picture for Baryons/Multi-Quarks}

Around 1980, the first application of lattice QCD Monte Carlo simulations 
\cite{C7980} was done by M.~Creutz for 
the static quark-antiquark (Q$\bar{\rm Q}$) potential. 
Since then, the study of inter-quark forces has been one of the important 
issues in lattice QCD \cite{R05}. 
Actually, in hadron physics, the inter-quark force can be regarded 
as an elementary quantity to connect the ``quark world'' to 
the ``hadron world'', and plays an important role to hadron properties. 

Around 2000, 
we performed the first accurate reliable lattice QCD study 
for the three-quark (3Q) potential, which is responsible to 
the baryon structure at the quark-gluon level. 
Furthermore, in 2005, we performed the first lattice QCD study 
for the multi-quark potentials, i.e., 4Q and 5Q potentials, 
which give essential information for the multi-quark hadron physics.
Note also that {\it the studies of 3Q and multi-quark potentials 
are directly related to the quark confinement properties 
in baryons and multi-quark hadrons.}

Here, we summarize our detailed studies of the inter-quark forces 
in three-quark/ multi-quark systems 
with SU(3)$_{\rm c}$ quenched lattice QCD \cite{TS0102,OST05,TS0304}.

In lattice QCD, the static Q$\bar {\rm Q}$ potential $V_{\rm Q\bar Q}(r)$ 
is calculated with the Wilson loop, and is well described with 
the Coulomb plus linear form as \cite{R05,TS0102} 
\begin{eqnarray}
V_{\rm Q \bar Q}(r)
=-\frac{A_{\rm Q\bar Q}}{r}+\sigma_{\rm Q \bar Q}r+C_{\rm Q\bar Q}.
\label{VQQ}
\end{eqnarray}
This physically means one-dimensional flux-tube formation 
between quark and antiquark, 
and this one-dimensional squeezing of color flux 
is shown in lattice QCD \cite{R05}.

For the color-singlet baryonic 3Q system, 
the 3Q potential can be calculated with the 3Q Wilson loop \cite{TS0102}.
For more than 300 different patterns of 3Q systems, 
we perform accurate calculation of the static 3Q potential $V_{\rm 3Q}$ 
in lattice QCD on various lattices:
($\beta$=5.7, $12^3\times 24$),
($\beta$=5.8, $16^3\times 32$), 
($\beta$=6.0, $16^3\times 32$),
($\beta=6.2$, $24^4$).
We find that $V_{\rm 3Q}$ is well 
described by the Coulomb plus Y-type linear potential (Y-Ansatz) 
\cite{TS0102,OST05,TS0304}, 
\begin{eqnarray}
V_{\rm 3Q}=-A_{\rm 3Q}\sum_{i<j}\frac1{|{\bf r}_i-{\bf r}_j|}+
\sigma_{\rm 3Q}L_{\rm min}+C_{\rm 3Q},
\label{V3Q}
\end{eqnarray}
where $L_{\rm min}$ is the minimal total length of 
of the color flux tube linking the quarks, i.e., 
the Y-shaped flux-tube length in most cases. 
As an example, we show in Fig.1(a) 
the 3Q confinement potential $V_{\rm 3Q}^{\rm conf}$, 
i.e., the 3Q potential subtracted by the Coulomb part, 
plotted against the Y-shaped flux-tube length $L_{\rm min}$.
At each $\beta$, clear linear correspondence is found between 
$V_{\rm 3Q}^{\rm conf}$ and $L_{\rm min}$, 
which indicates Y-Ansatz.
This physically means 
Y-shaped flux-tube formation among three quarks, and 
the Y-type flux-tube formation is actually observed in lattice QCD 
from the measurement of the action density  
in the spatially-fixed 3Q system \cite{Ichie03}, as shown in Fig.1(b).
Y-Ansatz has been also supported by many other recent studies in 
lattice QCD \cite{YAnsatz-lat} 
and theories including AdS/CFT \cite{YAnsatz-th}.

\begin{figure}[hb]
\includegraphics[height=3.9cm]{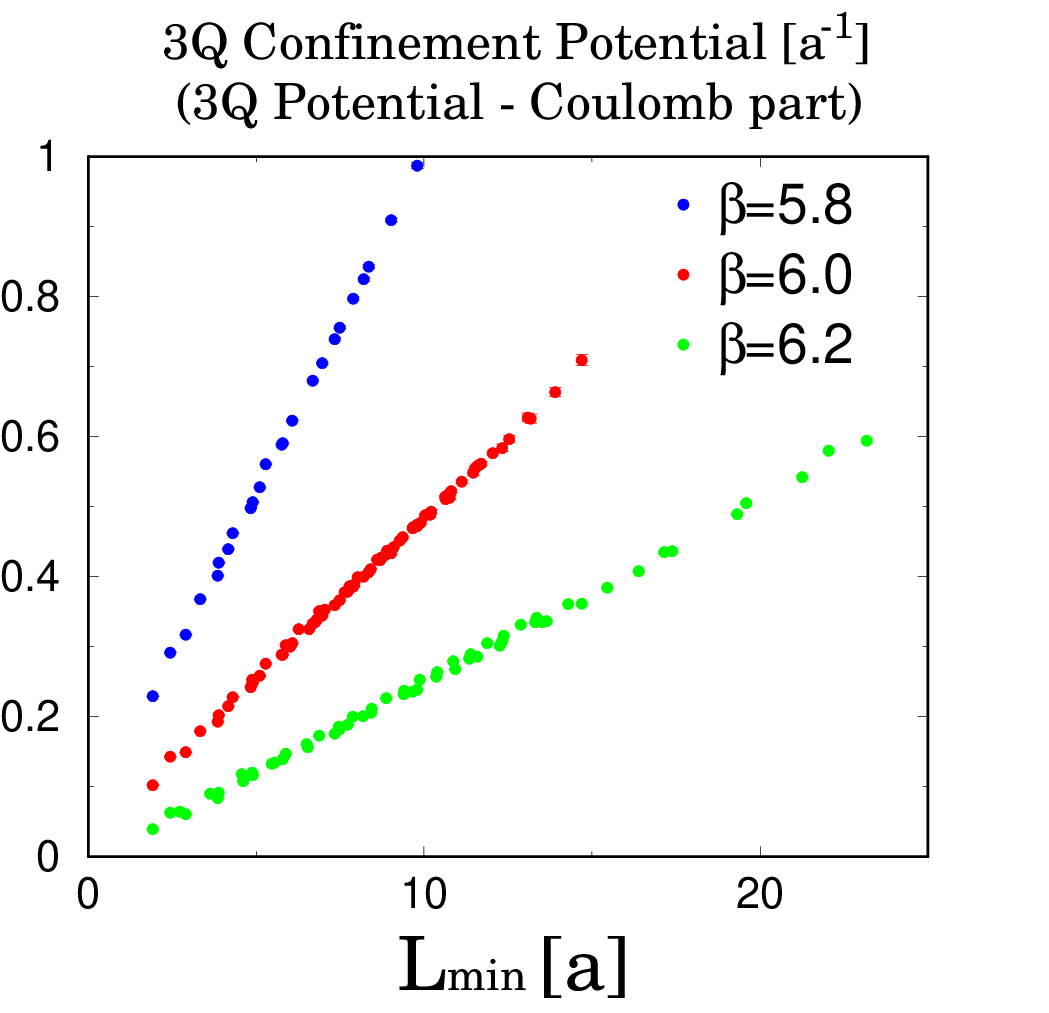}
\includegraphics[height=3.9cm]{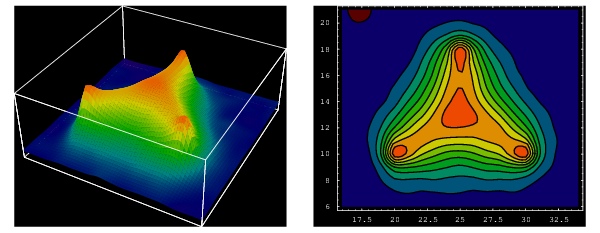}
\caption{
(a) The 3Q confinement potential $V_{\rm 3Q}^{\rm conf}$, 
i.e., the 3Q potential subtracted by the Coulomb part, 
plotted against $L_{\rm min}$ of Y-Ansatz
in lattice unit \cite{OST05}.
(b) The lattice QCD result for Y-type flux-tube formation 
in a spatially-fixed 3Q system \cite{Ichie03}.
The distance of each quark from the junction is about 0.5fm.
}
\end{figure}

After the experimental report of penta-quark candidates, 
lots of theoretical analyses for 
the exotic hadrons have been done or revisited. 
Recently, several charmed tetra-quark candidates 
such as X(3872) and Z$^+$(4430) have been experimentally discovered, 
and tetra-quark has been also investigated as an 
interesting object in quark-hadron physics.

Obviously, the quark-model calculation is one of the standard 
theoretical methods to investigate multi-quark systems.
In fact, the quark model calculation gives an outline of 
the properties of multi-quark hadrons and may 
predict new-type exotic hadrons theoretically. 
However, for such calculations, one needs 
the quark-model Hamiltonian for the multi-quark system. 
In particular, one has to know the quark confining potential in multi-quarks.
Then, we performed the first lattice QCD study of multi-quark potential.

We formulate the multi-quark Wilson loop, and accurately calculate 
the multi-quark potential in lattice QCD 
for about 200 different multi-quarks \cite{OST05}.
We find that the multi-quark potential 
obeys the OGE Coulomb plus 
string-theoretical linear potential \cite{OST05},
\begin{eqnarray}
V_{n{\rm Q}}=\frac{3}{2}A_{n{\rm Q}} \sum_{i<j}\frac{T^a_i T^a_j}
{|{\bf r}_i-{\bf r}_j|}+\sigma_{n{\rm Q}} L_{\rm min}+C_{n{\rm Q}}
~~~~ (n = 3,4,5,...),
\label{VnQ}
\end{eqnarray}
where the confinement potential is proportional to the minimal total 
length $L_{\rm min}$ of the color flux tube linking the quarks.
(Figure 2(a) and (b) are examples of the minimal-length flux-tube 
for the 4Q system.)
We find the {\it universality of the string tension}, 
$\sigma_{n{\rm Q}} \simeq \sigma_{\rm Q\bar Q}$, and 
the OGE result, $A_{n{\rm Q}} \simeq A_{\rm Q\bar Q}/2$, for $n=3,4,5$ 
\cite{TS0102,OST05}.

\begin{figure}[h]
\includegraphics[height=3.2cm]{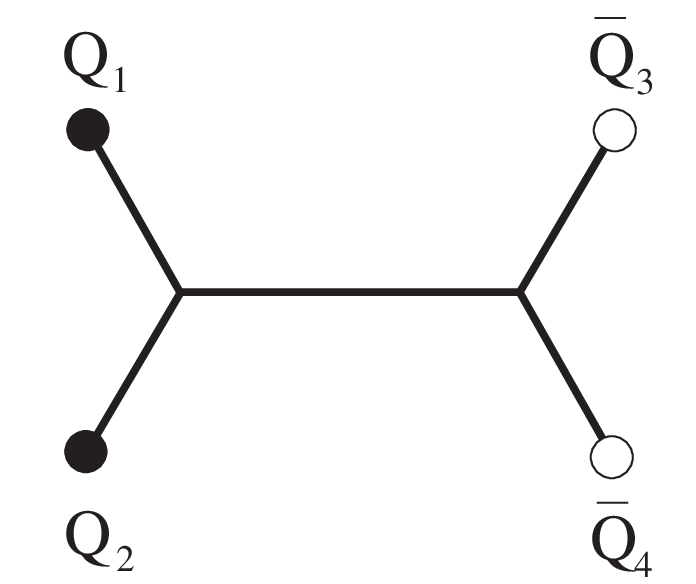}
\hspace{0.8cm}
\includegraphics[height=3.2cm]{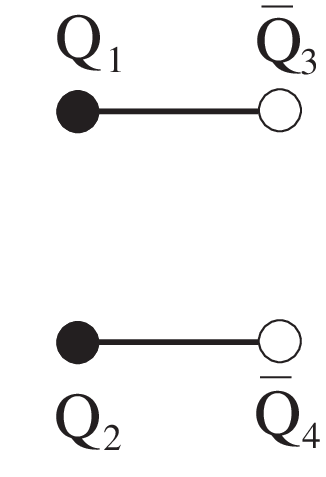}
\hspace{1cm}
\includegraphics[height=4.2cm]{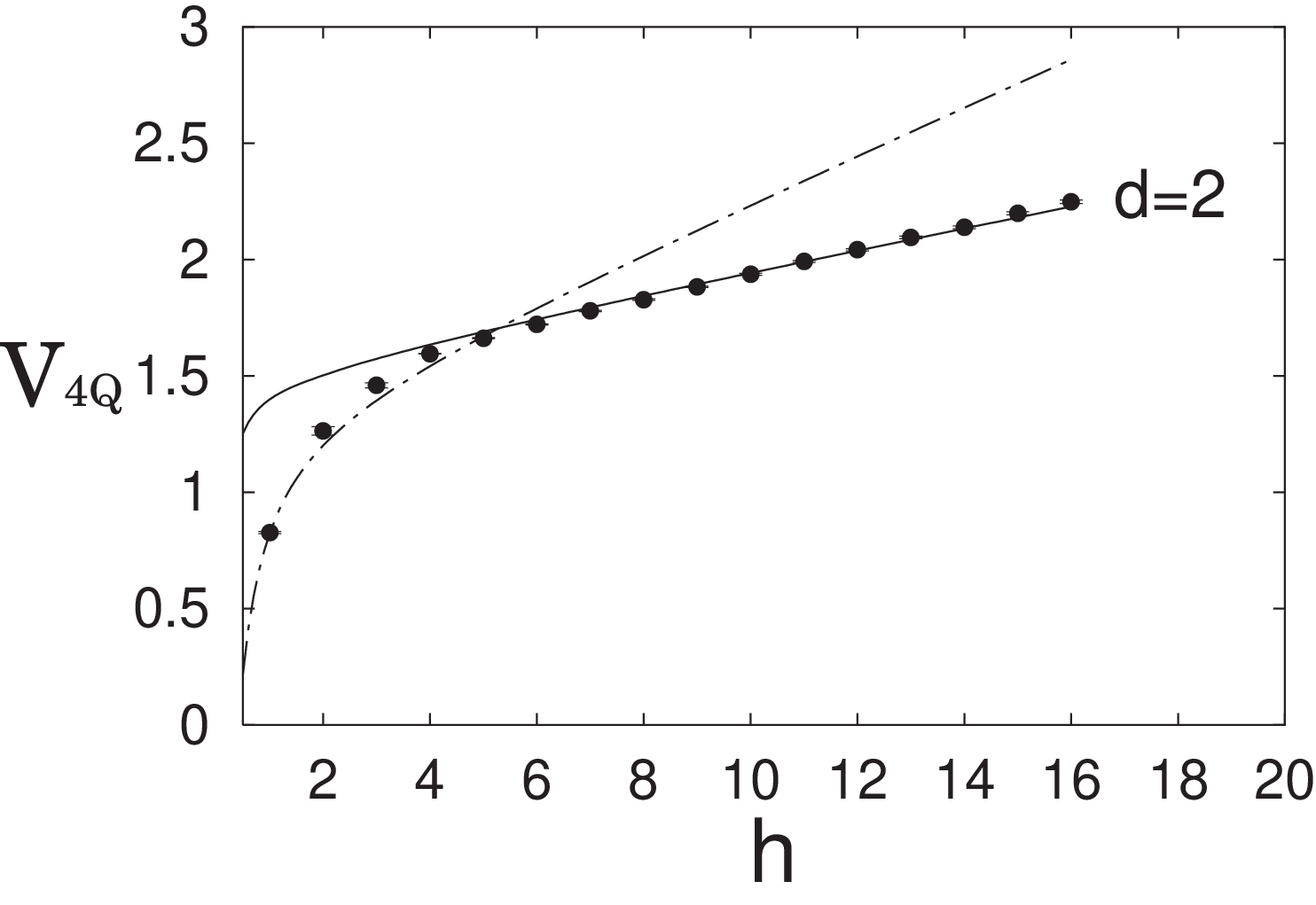}
\caption{
(a) A connected 4Q state and (b) a ``two-meson'' state.
(c) An example of the lattice QCD result of 
the 4Q potential $V_{\rm 4Q}$ 
for rectangular 4Q configurations in lattice unit \cite{OST05}.
The symbols denote the lattice QCD data. 
The solid (dotted-dashed) curve denotes connected 4Q (two-meson) state energy.
}
\end{figure}

For the multi-quark system, the topology of the flux-tube 
or the string linking quarks can be changed, according to the quark location. 
We also investigate the flux-tube recombination for 4Q systems 
\cite{OST05}. As the example, 
we show in Fig.2(c) the lattice QCD result of the 4Q potential $V_{\rm 4Q}$ 
for rectangular 4Q configurations as shown in Figs.2(a) and (b), 
with $d\equiv \overline{{\rm Q}_1{\rm Q}_2}/2$ and $h\equiv 
\overline{{\rm Q}_1{\rm Q}_3}$.
For large $h$, the lattice data obey the connected 4Q state energy.
For small $h$, the lattice data obey the ``two-meson'' state energy, i.e., 
$2V_{\rm Q\bar Q}(h)$. 
In fact, the 4Q potential $V_{\rm 4Q}$ is found to take 
the smaller energy of the connected 4Q state or the two-meson state 
\cite{OST05}.
In other words, we observe a clear lattice QCD evidence 
of the ``flip-flop'', i.e., the string recombination 
between the connected 4Q state and the two-meson state.

In this way, our lattice results physically indicates the validity of 
{\it infrared string picture}, not only for mesons 
but also for baryons and multi-quarks. 

Then, the multi-quark Hamiltonian based on lattice QCD is expressed as 
\begin{eqnarray}
H_{n{\rm Q}}= \sum_{i=1}^n (\hat {\bf p}_i^2+M_i^2)^{1/2}+ 
\sum_{i<j}^n V_{\rm OGE}({\bf r}_i-{\bf r}_j)+\sigma L_{\rm min}
~~~~ (n=3,4,5,...),
\end{eqnarray}
where $V_{\rm OGE}$ is the two-body OGE potential including 
the spin-dependent part, and $M_i$ the constituent quark mass.
The confinement part is described by infrared string picture.

As other recent developments of multi-quark/gluon potentials in lattice QCD, 
we studied the excited-state 3Q potential \cite{TS0304} and 
the heavy-heavy-light quark (QQq) potential \cite{YSI07}, 
and Cardoso-Bicudo studied the three-gluon static potential \cite{CB08}.

\section{Relevant Momentum of Gluon for Confinement}

Many theoretical physicists consider that  
confinement is brought by 
low-energy strong interaction of QCD.
However, no one has quantitatively showed 
the relevant energy region for confinement directly from QCD.
Here, the key question is 
{\it ``What is the important energy scale for confinement in QCD?''} 
Since confinement is mainly brought by gluon dynamics, 
this can be rewritten as 
{\it ``What is the relevant gluon momentum component 
responsible for confinement?''}

To answer this question, we formulate a new lattice method 
to extract relevant gluon momentum for each QCD phenomenon, 
based on four-dimensional (4D) discrete Fourier transformation \cite{YS0809}. 
For each gauge configuration, 
we select or remove gluon momentum components 
in lattice QCD by the following five steps.
\begin{enumerate}
\item
\noindent{\it Generation of link-variable}:
We generate gauge configurations on a $L^4$ lattice with the 
lattice spacing $a$ by the lattice-QCD Monte Carlo simulation 
under space-time periodic boundary conditions, and obtain 
coordinate-space link-variables $U_\mu(x)$. 
We here take the Landau gauge to suppress the artificial 
gauge fluctuation \cite{YS0809,ISI09}.
\item
\noindent{\it  Fourier transformation}:
By the 4D discrete Fourier transformation, 
we define the ``momentum-space link-variable'', 
${\tilde U}_{\mu}(p) \equiv \frac{1}{L^4}\sum_x 
U_{\mu}(x)\exp(i {\textstyle \sum_\nu} p_\nu x_\nu)$. 
The ``momentum-space lattice spacing'' is 
$a_p \equiv 2\pi/(La)$.

\item
\noindent{\it Cut in the momentum space}:
We introduce a cut (UV cut or IR cut) in the momentum space. 
Outside the cut, $\tilde U_\mu(p)$ is replaced 
by the free-field link-variable. 
We then obtain the momentum-space link-variable 
$\tilde U_\mu^\Lambda(p)$ with a cut. 
\item
\noindent{\it Inverse Fourier transformation}:
To return to coordinate space, we carry out the inverse Fourier transformation.
Since this $U'_\mu(x)$ is not an SU(3) matrix, we project it
onto an SU(3) element $U_\mu^\Lambda(x)$ by maximizing 
${\rm Re Tr} \{U^\Lambda_\mu(x)^\dagger U'_\mu(x)\}$.
Then, we get coordinate-space link-variable $U_\mu^\Lambda(x)$ with the cut.
\item
\noindent{\it Calculation}:
Using the cut link-variable $U_\mu^\Lambda(x)$ instead of $U_\mu(x)$, 
we calculate physical quantities as the expectation values
in the same way as ordinary lattice calculations.
This procedure is applicable to all the quantities in lattice QCD.
\end{enumerate}
With this method, we quantitatively determine 
the relevant gluon momentum component for confinement directly from QCD, 
through the analyses of the Q$\bar{\rm Q}$ potential \cite{YS0809}.
We perform quenched SU(3)$_{\rm c}$ lattice QCD calculations 
on $16^4$ lattice at $\beta$=5.7, 5.8 and 6.0.

\begin{figure}[h]
\includegraphics[scale=0.85]{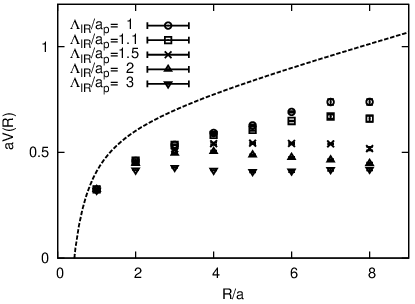}
\hspace{1cm}
\includegraphics[scale=0.85]{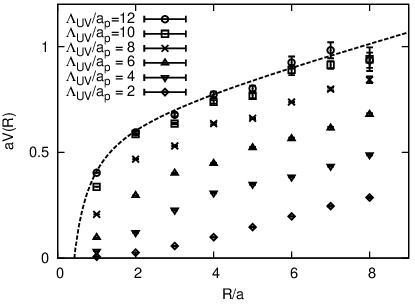}
\caption{
(a) The Q$\bar{\rm Q}$ potential $V(R)$ with the IR cut 
$\Lambda_{\rm IR}$ plotted against the inter-quark distance $R$.
(b) The Q$\bar{\rm Q}$ potential with the UV cut $\Lambda_{\rm UV}$.
The lattice QCD calculation is performed on $16^4$ lattice with $\beta =6.0$, 
{\it i.e.}, $a\simeq 0.10$fm and $a_p \equiv 2\pi/(La) \simeq 0.77$GeV.
The broken line is the original Q$\bar{\rm Q}$ potential.
}
\end{figure}

Figure 3 (a) and (b) show the Q$\bar{\rm Q}$ potential $V(R)$ 
with the IR cutoff $\Lambda_{\rm IR}$ 
and the UV cutoff $\Lambda_{\rm UV}$, respectively \cite{YS0809}.
Then, we obtain the following lattice-QCD results 
about the role of gluon momentum components on the inter-quark potential.
\begin{itemize}
\item
By the IR cutoff $\Lambda_{\rm IR}$, as shown in Fig.3(a), 
the Coulomb potential seems to be unchanged, 
but the confinement potential is largely reduced.
\item
By the UV cutoff $\Lambda_{\rm UV}$, as shown in Fig.3(b), 
the Coulomb potential is largely reduced, 
but the confinement potential is almost unchanged.
\end{itemize}

\begin{figure}[h]
\includegraphics[scale=0.9]{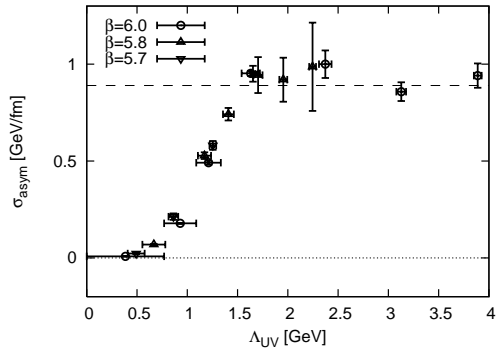}
\caption{
The $\Lambda_{\rm UV}$-dependence of the string tension 
$\sigma(\Lambda_{\rm UV})$ obtained from 
the asymptotic slope of the Q$\bar{\rm Q}$ potential 
$V(R)$ with the UV cutoff $\Lambda_{\rm UV}$. 
The lattice QCD calculations are done on $16^4$ lattice 
with $\beta$ =5.7, 5.8 and 6.0. 
The vertical error-bar is the statistical error, 
and the horizontal one is the range from 
the discrete momentum.
The broken line denotes the original value of 
the string tension $\sigma \simeq 0.89$GeV/fm.
}
\end{figure}

Figure 4 shows the $\Lambda_{\rm UV}$-dependence of the string tension 
$\sigma(\Lambda_{\rm UV})$ obtained from 
the asymptotic slope of the Q$\bar{\rm Q}$ potential 
$V(R)$ with the UV cutoff $\Lambda_{\rm UV}$ in the Landau gauge \cite{YS0809}.
(Similar results are obtained also in the Coulomb gauge.) 
As a remarkable fact, the string tension (confining force) 
is almost unchanged even after cutting off 
the high-momentum gluon component above 1.5GeV, 
while the string tension is significantly reduced 
when the UV cutoff is smaller than about 1.5GeV. 
We thus conclude that {\it quark confinement originates from 
the low-momentum gluon component below about 1.5GeV} \cite{YS0809}.

\section{Attempt of Linkage from QCD to Quark Model}

While QCD is the fundamental gauge theory of strong interaction, 
the quark model is also a successful model in describing hadrons. 
However, their relation is still unclear. 
We here consider two paths from QCD to the quark model, 
based on lattice-QCD results.

\subsection{Large Gluonic-Excitation Energy and Success of the Quark Model}

In the quark model, most low-lying hadrons can be well described 
only with quark degrees of freedom. 
{\it Why dynamical gluons do not appear in low-lying hadrons?}

We think that the absence of dynamical gluons 
is due to the large gluonic-excitation energy of 
about 1GeV for mesons \cite{JKM03} and baryons \cite{TS0304}.
(See Fig.5.) 
This is much larger than the quark-origin excitation such as 
spin-dependent interaction. 
Then, for low-lying hadrons, the gluonic excitation is absent,
and the system can be expressed only with quark degrees of freedom,
which gives a background of the success of the quark model. 
[In terms of the gluonic excitation, the hybrid hadron appears as 
a higher-excited state of about 1GeV, which explains the mass of 
Y(3940) as 4GeV $\simeq$ 1.5GeV~2 + 1GeV.]

\begin{figure}[h]
\includegraphics[height=4cm]{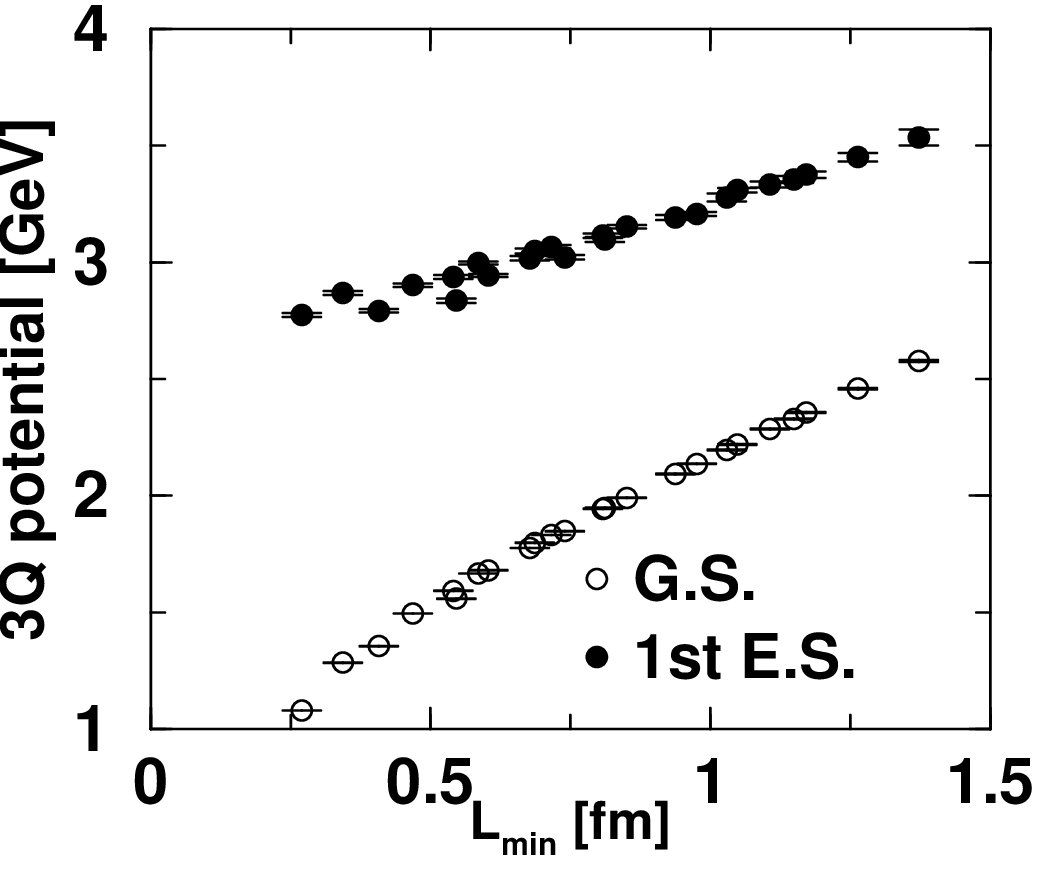}
\caption{ 
The 1st excited-state 3Q potential $V^{\rm e.s.}_{\rm 3Q}$ 
(filled circle) and the ground-state 3Q potential 
$V^{\rm g.s.}_{\rm 3Q}$ (open circle) 
v.s. $L_{\rm min}$ in lattice QCD 
at $\beta=5.8$ \cite{TS0304}. 
$\Delta E_{\rm 3Q} \equiv V^{\rm e.s.}_{\rm 3Q}-V^{\rm g.s.}_{\rm 3Q}$ 
is the gluonic-excitation energy.
}
\end{figure}

\subsection{What is the Gauge of QCD for the Quark Potential Model?}

The quark potential model is a nonrelativistic model 
with a potential instantaneously acting among quarks.
In this model, there are no dynamical gluons, 
and gluonic effects indirectly appear as the instantaneous 
inter-quark potential.
From the viewpoint of ``gauge'' in QCD, 
the quark model without dynamical gluons can be regarded as 
an effective theory of gauge-fixed QCD. 
Since the quark model has rotational and global color symmetries, 
such a gauge should keep them like Landau and Coulomb gauges. 

\begin{figure} [h]
\includegraphics[scale=0.45]{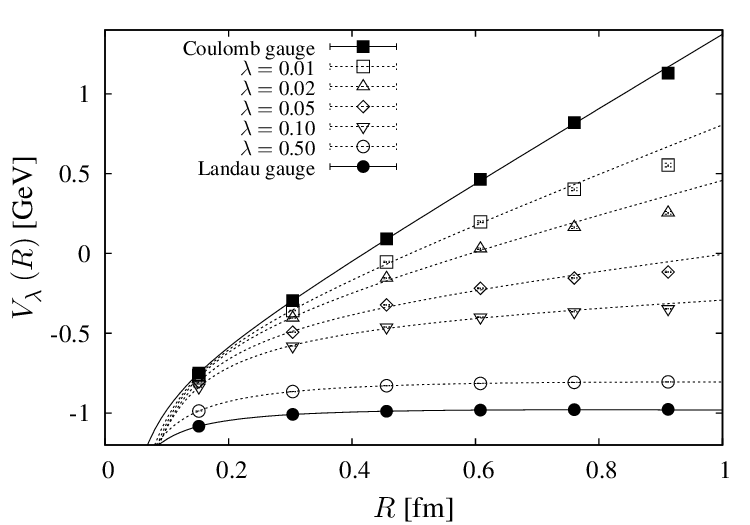}
\hspace{1cm}
\includegraphics[scale=0.45]{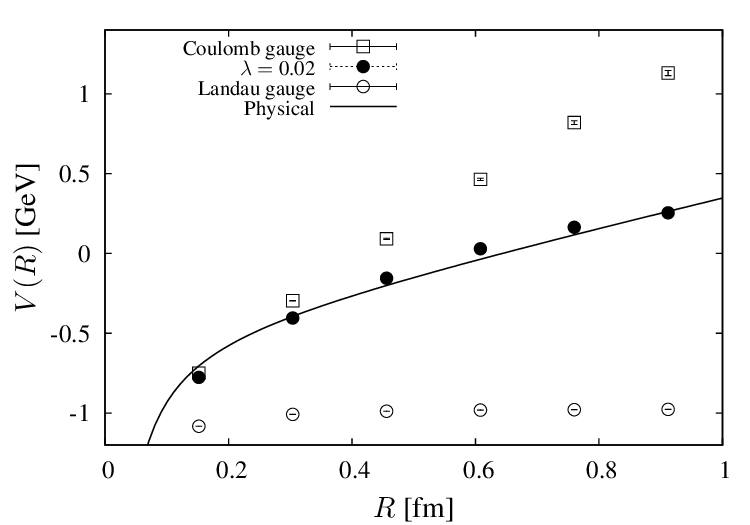}
\caption{
(a) ``Instantaneous potential'' $V_\lambda(R)$ in generalized Landau gauge
    for typical values of $\lambda$.
    The symbols are the lattice QCD results, and each curve 
    denotes the fit with the Coulomb plus linear form.
(b) Comparison of the instantaneous potential 
    $V_{\lambda}(R)$ at $\lambda = 0.02 (\simeq \lambda_C)$ (black dots) 
    with the physical static inter-quark potential (line).
We add the lattice data in Coulomb ($\lambda=0$) and 
Landau ($\lambda=1$) gauges.
}
\end{figure}

With a real parameter $\lambda \ge 0$, 
we define generalized Landau gauge ($\lambda$-gauge) \cite{IS11} of 
$\partial_i A_i+\lambda \partial_4 A_4=0$.
This gauge can connect continuously between 
the Landau ($\lambda$=1) and the Coulomb ($\lambda$=0) gauges. 
In lattice QCD, 
we investigate ``instantaneous potential'' 
$V_\lambda(R) \equiv - \frac{1}{a}\ln \langle {\rm Tr} 
[U_4^\dagger({\bf R},t)U_4({\bf 0},t)] \rangle$ 
in $\lambda$-gauge \cite{IS11}. 
Figure 6 shows the lattice result of 
the instantaneous potential \cite{IS11} : 
no linear part appears in the Landau gauge ($\lambda$=1); 
the Coulomb gauge ($\lambda$=0) shows overconfining 
with 2$\sim$3 times larger confining force \cite{GZ0304}, so that 
the ground state is conjectured to be the gluon-chain state \cite{GT02}.
In the $\lambda_C$-gauge with $\lambda_C \simeq 0.02$, 
the physical static inter-quark potential is approximately reproduced 
by the instantaneous potential \cite{IS11}, which seems to match 
the quark potential model.

\end{document}